\documentclass[sigconf,screen]{acmart}
\usepackage{amsmath}
\usepackage{enumitem}
\usepackage{multirow}
\usepackage{bm}
\usepackage{mathrsfs}
\usepackage{tcolorbox}
\usepackage{subfigure}
\AtBeginDocument{%
  \providecommand\BibTeX{{%
    \normalfont B\kern-0.5em{\scshape i\kern-0.25em b}\kern-0.8em\TeX}}}

  \usepackage{CJK}
  \usepackage{ulem}
\usepackage{booktabs}
\usepackage{array}

\newcommand\etal{{\it{et al.\ }}}

\newcommand{\finding}[2]{
\begin{tcolorbox}[width=\linewidth,boxrule=0pt,top=1pt, bottom=1pt, left=1pt,right=1pt, colback=gray!20,colframe=gray!20]
\textbf{Finding #1:} 
{#2}
\end{tcolorbox}}

\setcopyright{acmcopyright}
\acmPrice{15.00}
\acmDOI{10.1145/3540250.3549113}
\acmYear{2022}
\copyrightyear{2022}
\acmSubmissionID{fse22main-p307-p}
\acmISBN{978-1-4503-9413-0/22/11}
\acmConference[ESEC/FSE '22]{Proceedings of the 30th ACM Joint European Software Engineering Conference and Symposium on the Foundations of Software Engineering}{November 14--18, 2022}{Singapore, Singapore}
\acmBooktitle{Proceedings of the 30th ACM Joint European Software Engineering Conference and Symposium on the Foundations of Software Engineering (ESEC/FSE '22), November 14--18, 2022, Singapore, Singapore}
\begin{document}

\title{No More Fine-Tuning? An Experimental Evaluation of Prompt Tuning in Code Intelligence}

\author{Chaozheng Wang}
\affiliation{%
  \institution{Harbin Institute of Technology}
  \city{Shenzhen}
  \country{China}}
\email{wangchaozheng@stu.hit.edu.cn}

\author{Yuanhang Yang}
\affiliation{%
  \institution{Harbin Institute of Technology}
  \city{Shenzhen}
  \country{China}}
\email{ysngkil@gmail.com}

\author{Cuiyun Gao}
\authornote{Corresponding author. The author is also affiliated with Peng Cheng Laboratory and Guangdong Provincial Key Laboratory of Novel Security Intelligence Technologies. }
\affiliation{%
  \institution{Harbin Institute of Technology}
  \city{Shenzhen}
  \country{China}}
\email{gaocuiyun@hit.edu.cn}

\author{Yun Peng}
\affiliation{%
  \institution{The Chinese University of Hong Kong}
  \city{Hong Kong}
  \country{China}}
\email{ypeng@cse.cuhk.edu.hk}

\author{Hongyu Zhang}
\affiliation{%
  \institution{The University of Newcastle}
  \city{Newcastle}
  \country{Australia}}
\email{hongyu.zhang@newcastle.edu.au}

\author{Michael R. Lyu}
\affiliation{%
  \institution{The Chinese University of Hong Kong}
  \city{Hong Kong}
  \country{China}}
\email{lyu@cse.cuhk.edu.hk}
\begin{abstract}

Pre-trained models have been shown effective in many code intelligence tasks.
These models are pre-trained on large-scale unlabeled corpus and then fine-tuned in downstream tasks. However, as the inputs to pre-training and downstream tasks are in different forms, it is hard to fully explore the knowledge of pre-trained models. Besides, the performance of fine-tuning strongly relies on the amount of downstream data, while in practice, the scenarios with scarce data are common. Recent studies in the natural language processing (NLP) field show that prompt tuning, a new paradigm for tuning, alleviates the above issues and achieves promising results in various NLP tasks. In prompt tuning, the prompts inserted during tuning provide task-specific knowledge, which is especially beneficial for  tasks with relatively scarce data. In this paper, we empirically evaluate the usage and effect of prompt tuning in code intelligence tasks. We conduct prompt tuning on popular pre-trained models CodeBERT and CodeT5 and experiment with three code intelligence tasks including defect prediction, code summarization, and code translation. Our experimental results show that prompt tuning consistently outperforms fine-tuning in all three tasks. In addition, prompt tuning shows great potential in low-resource scenarios, e.g., improving the BLEU scores of fine-tuning by more than 26\% on average for code summarization. Our results suggest that instead of fine-tuning, we could adapt prompt tuning for code intelligence tasks to achieve better performance, especially when lacking task-specific data.
\end{abstract}

\ccsdesc[500]{Software and its engineering}
\ccsdesc[500]{Software and its engineering~Software development techniques}
\keywords{code intelligence, prompt tuning, empirical study}
 \maketitle
\section{Introduction}\label{sec:intro}
\begin{figure*}
    \centering
    \includegraphics[width=0.9\textwidth]{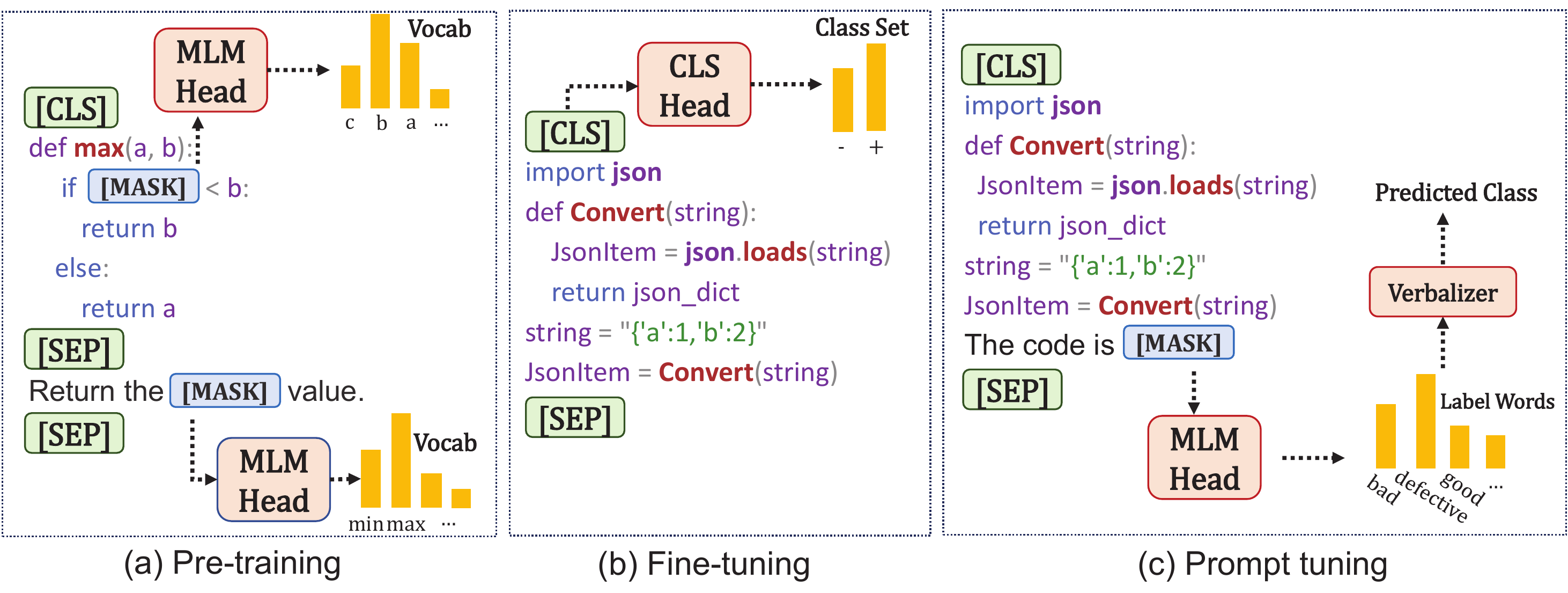}
    \caption{Illustration on the process of pre-training, fine-tuning, and prompt tuning on defect detection task. $[CLS]$ and $[SEP]$ denote two special tokens in pre-trained models.}
    \label{fig:intro}
\end{figure*}


Code intelligence leverages machine learning, especially deep learning (DL) techniques to mine knowledge from large-scale code corpus and build intelligent models for improving the productivity of computer programming. 
The state-of-the-art DL-based approaches to code intelligence exploit the \textit{pre-training and finetuning} paradigm~\cite{feng2020codebert, wang2021codet5, ahmad2021unified,kanade2020learning, guo2020graphcodebert}, in which language models are first pre-trained on a large {unlabeled} text corpora  
and then finetuned on downstream tasks.
For instance, Feng \etal~\cite{feng2020codebert} propose CodeBERT, a pre-trained language model for source code, {which leverages both texts and code in the pre-training process.}
{To facilitate generation tasks for source code, }
Wang \etal~\cite{wang2021codet5} 
propose a pre-trained sequence-to-sequence model
named CodeT5.
{These pre-trained source code models achieve significant improvement over previous approaches. }

However, there exist
gaps between 
the pre-training and fine-tuning process of these pre-trained models.
As shown in Figure~\ref{fig:intro}(a), pre-training models such as CodeBERT~\cite{feng2020codebert} and CodeT5~\cite{wang2021codet5} are generally
pre-trained using the Masked Language Modeling (MLM) objective. The input to MLM is a mixture of code snippets and natural language texts, and the models are trained to predict randomly-masked input tokens.
However, when models are fine-tuned into the downstream tasks, e.g. defect detection, the input involves only source code and the training objective changes to a classification problem. As shown in Figure~\ref{fig:intro}(b), the pre-trained model represents each input code snippet using a classification head (CLS Head) and fine-tunes the CLS head based on a task-specific dataset.
The inconsistent inputs and objectives between pre-training and fine-tuning render the knowledge of pre-trained models hard to be fully explored, leading to sub-optimal results for downstream tasks. Besides, the performance of fine-tuning largely depends on the scale of downstream data \cite{lester2021power, han2021ptr, gu2021ppt, zhang2021differentiable}.

Recently, prompt tuning \cite{li2021prefix, schick2021exploiting, han2021ptr, liu2021p, lester2021power} is proposed to mitigate the above issues of fine-tuning.
Figure \ref{fig:intro}(c) illustrates the concept of prompt tuning. Instead of only involving source code as input, 
prompt tuning firstly rewrites the input by adding a natural language prompt such as $``The\ code\ is\ [MASK]"$ at the end of the code snippet, and then let the model predict the masked token $[MASK]$. There is also a verbalizer \cite{schick2021exploiting, han2021ptr} that maps the tokens predicted by the model to the class. By adding a prompt and verbalizer, prompt tuning reformulates the classification problem into an MLM problem, aligning the objective with the
pre-training stage. This alignment unleashes the hidden power stored in the pre-trained models. Besides, the inserted natural language prompt can involve task-specific
knowledge to facilitate the adaption
to downstream tasks \cite{li2021prefix, schick2021exploiting, shen2021partial, guo2019spottune}. 

Inspired by the success of prompt tuning in the NLP field, we would like to investigate if prompt tuning is effective for code intelligence tasks, which to our best knowledge still remains unexplored.
In this paper, we conduct an experimental evaluation on the effectiveness of prompt tuning on three popular code intelligence tasks: defect detection, code translation, and code summarization. We mainly investigate the following three research questions (RQs):
\begin{itemize}
    \item How effective is the prompt tuning in solving code intelligence tasks?
    \item How capable is prompt tuning to handle data scarcity scenarios?
    \item How different prompt templates affect the performance of prompt tuning? 
\end{itemize}
To answer the first RQ, we apply prompt tuning to the three  code intelligence tasks. To answer the second RQ, we evaluate prompt tuning in data scarcity scenarios from two aspects, including low-resource settings and cross-domain settings. To answer the third RQ, we comprehensively study the influence of different prompt templates and verbalizers on model performance.

Based on the experiment results, we find that prompt tuning brings non-trivial improvement to the the performance of downstream code intelligence tasks, including both classification and generation tasks. Furthermore, prompt tuning can significantly outperform conventional fine-tuning, especially when the training data are scarce. 



{The major contributions of this paper are as follows:}
\begin{enumerate}
    \item To the best of our knowledge, this paper serves as the first study on the performance of prompt tuning for code intelligence tasks.
    \item We explore how different prompts can affect the performance of prompt tuning on code intelligence tasks.
    \item {We discuss the implications of our findings and suggest further research on the usage of prompt tuning.} 
\end{enumerate}

\section{BACKGROUND}\label{sec:prompt}
\begin{figure}
    \centering
    \includegraphics[width=0.45\textwidth]{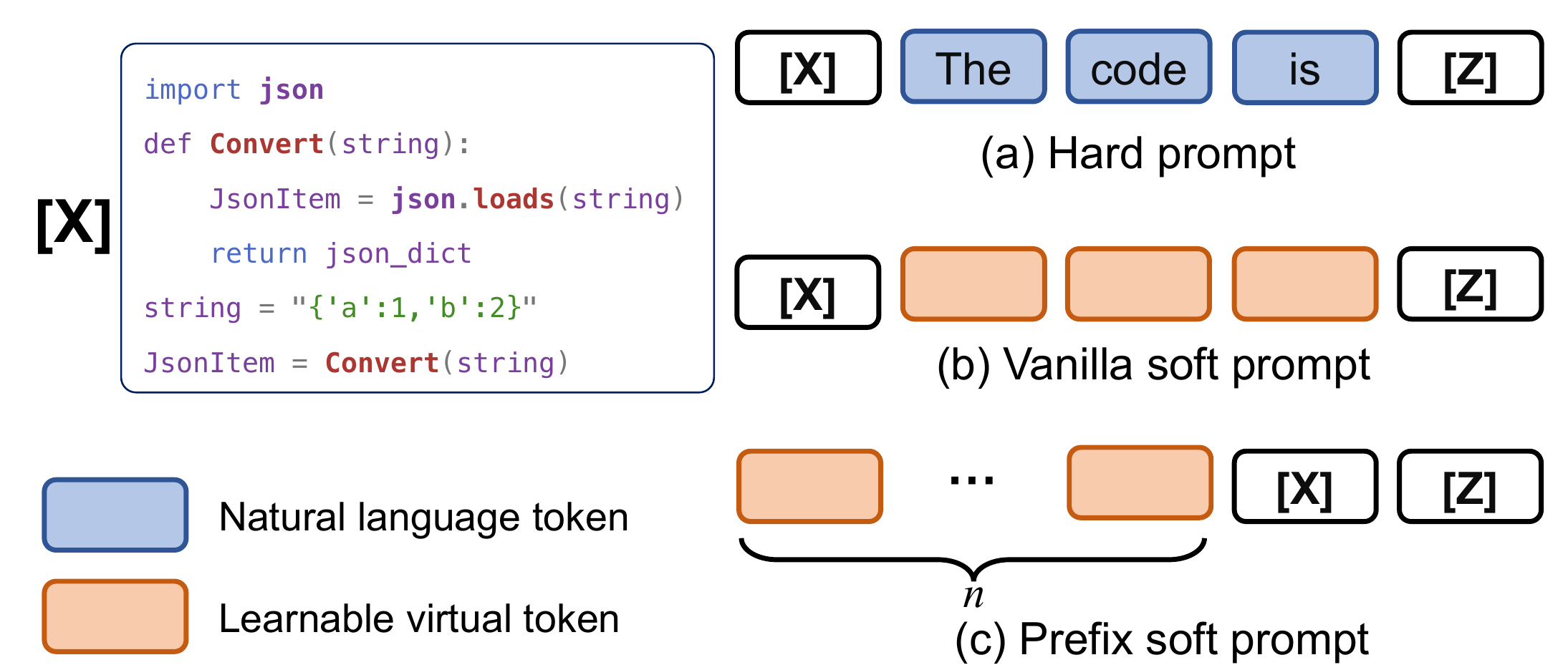}
    \caption{Illustration on the different types of prompt, where $[X]$ and $[Z]$ indicate the input slot and answer slot, respectively. Both vanilla soft prompt (b) and prefix soft prompt (c) belong to soft prompt.}
    \label{fig:prompt}
\end{figure}
\subsection{Fine-tuning}
Fine-tuning a pre-trained model for downstream tasks~\cite{liu2016recurrent,peters2018deep, devlin2018bert} is a prevalent paradigm in the NLP field. Fine-tuning aims at exploiting the knowledge learned by pre-trained models without learning from scratch and can be regarded as a way of applying transfer learning~\cite{pan2009survey}. To adapt pre-trained models into downstream tasks, fine-tuning trains the model in a supervised way. 
Specifically, given a dataset which consists of task-specific samples $X$ and corresponding labels $Y$, fine-tuning aims to find a set of parameters $\theta$ for the pre-trained
model, that 
$\theta = \mathop{\arg\min}\limits_{\theta} P(Y | X ;\theta)$.
\subsection{Prompt Tuning}
The intuition of prompt tuning is to convert the training objective of downstream tasks into a similar form as the pre-training stage, i.e., the MLM objective \cite{devlin2018bert, liu2019roberta, feng2020codebert}.
As shown in Figure~\ref{fig:intro}(c), prompt tuning aims at predicting masked tokens in the input. It also modifies the model input by adding a natural language prompt, enabling the input format identical to the pre-training stage.

Specifically, prompt tuning employs a prompt template $f_{prompt}(x)$ to reconstruct the original input $\bm{x}$, producing new input $\bm{x}'$. As illustrated in Figure~\ref{fig:prefix_sum}, the prompt template can involve two types of reserved slots in, i.e., input slot $[X]$ and answer slot $[Z]$. The input slot $[X]$ is reserved to be filled with original input text, and the answer slot $[Z]$ is to be filled by predicted labels such as \textit{defective}. For the example shown in Figure~\ref{fig:intro}, prompt tuning outputs the final predicted class by a verbalizer \cite{han2021ptr, schick2021exploiting}. The verbalizer, denoted as $\mathcal{V}$, is an injective function which maps each predicted label word to a class in the target class set $Y$: 
\begin{equation}
\mathcal{V}: W \rightarrow Y
\end{equation}
\noindent where $W$ indicates the label word set. For the example in Figure~\ref{fig:intro} (c), the label word set $W$ includes ``\textit{[bad, defective]}'' for buggy code snippets and ``\textit{[perfect, clean]}'' for the others. The class set $Y$ contains ``$+$'' and ``$-$'' for indicating defective and clean code, respectively. In the example, the verbalizer maps the label with highest probability ``\textit{defective}'' into the target class ``$+$'' in the class set.

According to the flexibility of the inserted prompt, prompt tuning techniques can be categorized into two types: hard prompt and soft prompt. We elaborated on the details of each prompt type in the following.


\subsubsection{Hard Prompt} The \textit{hard prompt} \cite{schick2021exploiting, han2021ptr, gu2021ppt} is a technique that modifies the model input by adding fixed natural language instruction (prompts). It aims to elicit task-specific knowledge learned during pre-training for the tuning stage. Hard prompt is also known as \textit{discrete prompt} since each token in the prompts is meaningful and understandable \cite{gu2021ppt, liu2021pre}. 
For instance, in the defect detection task, by appending ``The code is [Z]." to the input code, 
the task objective becomes predicting the label word
at the answer slot
$[Z]$, such as ``\textit{defective}" or ``\textit{clean}". The designed prompt template for defect prediction task can be formulated as:
\begin{equation}
    f_{prompt}(x) = ``[X] \ The \ code \ is \ [Z]"
\end{equation}
where $[X]$ denotes the input code.
Although hard prompt has shown promising performance in previous work, the template design and the verbalizer choices are challenging. For example, the prompt template $f_{prompt}(x)$ can also be designed as $``[X] \ It \ is \ [Z]"$, where the label words in the verbalizer involve ``\textit{bad}'' and ``\textit{perfect}''. 

\subsubsection{Soft Prompt}\label{sec:soft}
The \textit{Soft prompt} \cite{han2021ptr, li2021prefix, tsimpoukelli2021multimodal}, as the name implies, 
is an alternative to hard prompt.
Different from hard prompt, the tokens in the soft prompt template are not fixed discrete words of a natural language. Instead, these tokens are continuous vectors which can be learnt during the tuning stage.
They are also called \textit{virtual tokens} because they are not human-interpretable. Soft prompt is proposed to
alleviate the burden of manually selecting prompt template in hard prompt.
There are two kinds of soft prompt, denoted as \textit{vanilla soft prompt} and \textit{prefix soft prompt}, respectively. 

\textit{Vanilla soft prompt}, as depicted in Figure~\ref{fig:prompt}(b), can be obtained by simply replacing the hard prompt token with a virtual one, denoted as $[SOFT]$, such as:
\begin{equation}
    f_{prompt}(x) = ``[X] \ [SOFT] \ [SOFT] \ [SOFT] \ [Z]"
\end{equation}
The embedding of virtual tokens are optimized during tuning stage.

\textit{Prefix soft prompt} prepends several virtual tokens to the original input, as shown in  Figure~\ref{fig:prompt}(c).
It can generate comparable performance with the
vanilla soft prompt and hard prompt. 
\begin{equation}\label{equ:prefix}
    f_{prompt}(x) = ``[SOFT] * n \ [X] \ [Z]"
    \end{equation}
\noindent where $n$ indicates the number of virtual tokens.



\section{Experimental Evaluation}\label{sec:Methodology}

\subsection{Research Questions}\label{sec:RQs}
We aim at answering the following research questions through an extensive experimental evaluation: 
\begin{enumerate}[label=\bfseries RQ\arabic*:,leftmargin=.5in]
    \item How effective is the prompt tuning in solving code intelligence tasks?
    \item How capable is prompt tuning to handle data scarcity scenarios?
    \item How different prompt templates affect the performance of prompt tuning?
\end{enumerate}

We design RQ1 to verify our hypothesis that prompt tuning, which aligns the training objectives with the pre-training stage, is more effective than fine-tuning for the downstream code intelligence tasks. RQ2 aims at investigating whether prompt tuning embodies advantage in data scarcity scenarios including low-resource and cross-domain settings. In RQ3, we aim at exploring the impact of different prompt templates, such as varying prompt types and selection of label words, on the  performance of downstream tasks.


\begin{table}[ht]
    \centering
    \caption{Statistics of the datasets used in this paper.
    }
    \begin{tabular}{c|c|rrr}
    \toprule
      \multirow{2}{*}{Tasks}  & \multirow{2}{*}{Datasets} & Training  & Val.  & Test  \\
      & & Set & Set & Set\\
      
       \midrule
       Defect Detection & Defect & 21,854 & 2,732 & 2,732 \\
       
       \midrule
       & Ruby &48,791 &2,209 & 2,279 \\
       & JavaScript &123,889 &8,253 & 6,483 \\
        
           
         Code & Go & 317,832& 14,242&14,291 \\
     Summarization & Python & 409,230& 22,906& 22,104\\
        & Java &454,451& 15,053& 26,717  \\
        & PHP & 523,712& 26,015& 28,391 \\
        \midrule
       {Code Translation} & Translation &{10,300} & {500} & {1,000} \\
    \bottomrule
       
    \end{tabular}
    
    \label{tab:dataset}
\end{table}

\subsection{Code Intelligence Tasks with Prompt Tuning}\label{sec:Methodology}

     
    

To evaluate the prompt tuning  technique on source code, we adopt
three downstream code intelligence tasks, namely defect detection, code summarization, and code translation. 
We describe the detail of pre-trained models and prompt template of each task in the following.
\subsubsection{Pre-trained Models}
We choose CodeBERT~\cite{feng2020codebert} and CodeT5 \cite{wang2021codet5} as the studied pre-trained models, since they are the most widely-used model and state-of-the-art model for source code, respectively.

\textbf{CodeBERT} \cite{feng2020codebert} is an encoder-only model which is realized based on RoBERTa \cite{liu2019roberta}. CodeBERT is pre-trained on CodeSearchNet~\cite{husain2019codesearchnet}. It is able to encode both source code and natural language text. CodeBERT has 125 million parameters. 

\textbf{CodeT5} \cite{wang2021codet5}, a variant of text to text transfer Transformer \cite{raffel2019exploring}, is the state-of-the-art model for code intelligence tasks. It regards all the tasks as sequence to sequence paradigm with different task specific prefixes. It can solve both code understanding and code generation tasks. Code-T5 is pre-trained on a larger dataset including CodeSearchNet~\cite{husain2019codesearchnet} and an additional C/C\# language corpus collected by the authors. CodeT5 is 
classified into two versions: CodeT5-small and CodeT5-base, according to their sizes. The numbers of parameters in CodeT5-small and CodeT5-base are 60 million and 220 million, respectively.



 
\subsubsection{Defect Detection}
Given a code snippet,
defect detection~\cite{zhou2019devign, li2021vuldeelocator} aims to identify whether 
it is defect prone,
such as memory leakage and DoS attack. 
The task is defined as a binary classification task in training CodeBERT and a generation task in training CodeT5~\cite{wang2021codet5, raffel2019exploring}. 

For \textit{hard prompt}: As shown in Figure~\ref{fig:intro}(c), with prompt tuning,
models predict the probability distribution over the label words. A verbalizer $\mathcal{V}$ maps the label word with highest probability to the predicted class. One cloze-style template $f_{prompt}(\cdot )$ with an input slot $[X]$ and an answer slot $[Z]$ is designed as below:

\begin{equation}
\begin{aligned} 
    f_{prompt}(x) &= ``The\; code\; [X] \;is\; [Z]" \\
    \mathcal{V} &= \left\{
    \begin{aligned}
        & \bm{+}: &[defective, bad] \\
        & \bm{-}: &[clean, perfect]
    \end{aligned}
    \right.
\end{aligned}
\end{equation}
where the left and right sides of $:$ indicate the predicted class and corresponding label words. To study the impact of different prompts, we also design other prompt templates including $``[X]\; It\; is\; [Z]"$, $``[X]\; The \;code\; is\; [Z]"$, $``[X]\; The\; code\; is\; defective\; [Z]"$ and $``A \;[Z]\; code\; [X]"$.

For \textit{vanilla soft prompt}: For facilitating the comparison of hard prompt and vanilla soft prompt, we simply replace the natural language tokens in the hard prompt templates with virtual tokens for generating vanilla soft prompts. For example, $``[X] [SOFT] [SOFT] [Z]"$ is the vanilla soft prompt version of $``[X]\; It\; is\; [Z]"$.


For \textit{prefix soft prompt}: We design the prefix soft prompt by appending a learnable prefix prompt according to Equation (\ref{equ:prefix}).


\subsubsection{Code summarization}\label{sec:sum}
Given a code snippet, the code summarization task aims to generate a natural language comment to summarize the functionality of the code. We only utilize CodeT5 in this task because CodeBERT does not have a decoder to generate comments.


For \textit{hard prompt}: We append the natural language instruction of the task to the input code, so the template can be:
\begin{equation}
\begin{aligned}
    f_{prompt}(x) &= ``\;Generate\;comment\;for\;[LANG]\;[X]\;[Z]" \\
\end{aligned}
\end{equation}
where $[LANG]$, $[X]$, and $[Z]$ denote the slot of programming language type, input slot, and the generated answer slot. The natural language instruction ``\textit{Generate comment for}'' is manually pre-defined for adjusting the generation behavior of CodeT5. We also design other prompt templates for experimentation including \textit{Summarize [LANG]}. Note that there is \textit{no verbalizer} for the generation task.

For the \textit{vanilla soft prompt} and \textit{prefix soft prompt}: They are designed in the same way as the defect detection task. For example, we replace the natural language tokens in the hard prompt templates into virtual tokens for generating vanilla soft prompts. The prefix soft prompts are defined according to Equation (\ref{equ:prefix}).

\subsubsection{Code Translation}
Code translation aims to migrate legacy software from one programming language
to another one. The \textit{vanilla soft prompt} and \textit{prefix soft prompt} are designed similar to the above two tasks, so we only describe about the \textit{hard prompt} for the task.
For \textit{hard prompt}, we design the template by appending task-specific instruction:
\begin{equation}
\begin{aligned}
    f_{prompt}(x) &= `` Translate\; [X] \; to \; [LANG] \; [Z]" \\
\end{aligned}
\end{equation}

The template explains that the model is translating the input code $[X]$ in one programming language to the code $[Z]$ in another programming language $[LANG]$.




\subsection{Evaluation Datasets}
To empirically evaluate the performance of prompt tuning for source code, we choose the datasets for the three tasks from the popular CodeXGLUE benchmark\footnote{https://github.com/microsoft/CodeXGLUE} \cite{lu2021codexglue}. 


\subsubsection{Defect Detection} The dataset 
is provided by Zhou \etal \cite{zhou2019devign}. It contains 27K+ C code snippets from two open-source projects QEMU and FFmpeg, and 45.0\% of the entries are defective.

\subsubsection{Code Summarization}   
We use the same dataset as the CodeT5 work~\cite{wang2021codet5}. The dataset is from CodeSearchNet \cite{husain2019codesearchnet}, which contains thousands of code snippet and natural language description pairs for six programming languages including Python, Java, JavaScript, Ruby, Go and PHP.


\subsubsection{Code translation} The dataset is provided by Lu \etal ~\cite{lu2021codexglue}, and is collected from four public repositories (including Lucene, POI, JGit and Antlr).
Given a piece of Java (C\#) code, the task is to translate the code into the corresponding C\# (Java) version.



\subsection{Evaluation Metrics}

\subsubsection{Defect Detection:}
For the defect detection task, following \cite{wang2021codet5}, we use \textit{Accuracy} as the evaluation metric. The metric is to measure the ability of model to identify insecure source code, defined as:
\begin{equation}\label{equ:acc}
    ACC = \frac{\sum^{|D|}_{i=1}1(y_i==\hat{y}_i)}{|D|}
\end{equation}
where $D$ is the dataset and $|D|$ denotes its size. The symbol $y_i$ and $\hat{y}_i$ indicate the ground truth label and predicted label, respectively. The $1(x)$ function returns 1 if $x$ is True and otherwise returns 0.

\subsubsection{Code Summarization:}
Following previous work \cite{wang2021codet5, feng2020codebert}, we use Bilingual Evaluation Understudy (BLEU) score to evaluate the quality of generated comments. The idea of BLEU is that the closer the generated text is to the result of ground truth text, the higher the generation quality. The metric is defined as below: 

\begin{equation}
       BP = \left\{
    \begin{aligned}
        &1 & if \; c>r \\
        &e^{1-r/c} &if c\leq r \\
    \end{aligned}
    \right.
\end{equation}

\begin{equation}
    BLEU = BP \cdot exp\Big({\sum^{N}_{n=1}{w}_{n}\log{p}_{n}}\Big)
\end{equation}
\noindent where $p_{n}$ means the modified n-gram precision and ${w}_{n}$ is the weight. $BP$ represents the brevity penalty, and $c$ and $r$ indicate the lengths of generated comment and target comment length, respectively. In our experiments, we choose smoothed BLEU-4 score, i.e., $n=4$,
for evaluating the generation tasks following previous work \cite{feng2020codebert, wang2021codet5}.

\subsubsection{Code Translation:}
To better measure the quality of generated code snippets, besides BLEU score, another two metrics including Accuracy and CodeBLEU \cite{ren2020codebleu} are used following
\cite{lu2021codexglue, wang2021codet5}. The computation of Accuracy is the same as Equ. (\ref{equ:acc}), which is the most strict metric. 

CodeBLEU parses the generated code, and takes both the code structure and semantics into account for measuring the similarity between the generated code and the code in ground truth. Its computation
consists of four components including n-gram matching score ($BLEU$), weighted n-gram matching score $weighted\_BLEU$, syntactic AST matching score $AST\_Score$ and semantic data flow matching score $DF\_Score$:
\begin{equation}
\begin{aligned}
    CodeBLEU &= \alpha * BLEU + \beta * weighted\_BLEU \\
             &+ \gamma * AST\_Score + \delta * DF\_Score
\end{aligned}
\end{equation}
where $\alpha, \beta, \gamma, \delta$ are weights for each component. Following \cite{wang2021codet5, lu2021codexglue}, they are all set as 0.25.

\begin{table}[]
    \centering
    \caption{Hyperparameter settings}
    \scalebox{0.92}{\begin{tabular}{c|c||c|c}
    \toprule
       Hyperparameter  & Value &  Hyperparameter  & Value\\
     \midrule
      Optimizer & AdamW\cite{loshchilov2018decoupled} & Warm up steps   & 10\% \\
      Learning rate & 5e-5 & Training batch size & 64 \\
      LR scheduler& Linear& Validation batch size & 64 \\
        Beam size & 10 &  Adam epsilon & 1e-8 \\
      Max. gradient norm & 1.0 & &\\
      \bottomrule
    \end{tabular}
    }
    
    \label{tab:param}
\end{table}

\subsection{Implementation Details}
\subsubsection{Experimental Setup}
All the pre-trained models and corresponding tokenizer
in our experimentation are loaded from the official repository
Huggingface\footnote{https://huggingface.co/models}. The overall framework is Pytorch\footnote{https://pytorch.org/}. Our implementation of prompt is based on OpenPrompt \cite{ding2021openprompt}. We 
use the generic training strategy and parameter settings following the official implementation of CodeT5 \cite{wang2021codet5}, with details shown in Table \ref{tab:param}.

Specifically, for the defect detection task, we train CodeBERT and CodeT5 for 5 and 15 epochs, respectively. For CodeT5 model, we set the maximum  source length and target length as 512 and 3, respectively. 
For the code summarization task, because CodeBERT is an encoder-only architecture model, we focus on evaluating prompt tuning on CodeT5.
We train CodeT5 for 20 epochs. The maximum lengths of source text and target text are defined as 256 and 128.
For the code translation tasks, we train the CodeT5 models for 50 epochs. The maximum length of source text and target text are set as 256 and 256, respectively. 

For parameter configuration in fine-tuning, we use the configurations provided by the original work \cite{feng2020codebert, wang2021codet5}, which were already well adjusted. For a fair comparison, we use the same parameter configurations when implementing prompt tuning.

All the experiments are run on a server with 4 * Nvidia Tesla V100 and each one has 32GB graphic memory.

\subsubsection{Fine-tuning Baselines}
We fine-tuned CodeBERT and CodeT5 on the three code intelligence tasks.
Specifically, we fine-tune CodeBERT only for the defect detection and CodeT5 for all the three tasks. For CodeBERT, we use the first output token (the [CLS] token) as the sentence embedding and feed it into a feed-forward network (FFN) to generate predictions. For CodeT5, all the tasks are treated as generation tasks. It takes either code or natural language sentences as input and generate target texts.


\begin{table}[ht]
    \centering
    \caption{Classification accuracy on defect detection.}
    \begin{tabular}{cc|c}
    \toprule
       \multicolumn{2}{c|}{Methods}  & Accuracy \\
      \midrule
    \multirow{2}{*}{CodeBERT} & Fine-tuning   & 62.12 \\
      & Prompt tuning & \textbf{64.17} \\
      
      \midrule
     \multirow{2}{*}{CodeT5-small} & Fine-tuning    & 62.96 \\
      & Prompt tuning & \textbf{63.91} \\
      
      \hline
    \multirow{2}{*}{CodeT5-base} & Fine-tuning    & 65.00 \\
      & Prompt tuning & \textbf{65.82} \\
      \bottomrule
    \end{tabular}
    \label{tab:defect}
\end{table}

\begin{table*}[ht]
    \centering
    \caption{Results (BLEU-4 scores) of the CodeT5 model on code summarization task. }
    \vspace{-6pt}
    \begin{tabular}{cc|cccccc|c}
    \toprule
    \multicolumn{2}{c}{Methods}     & Ruby & JavaScript & Go & Python & Java & PHP & Overall \\

    \midrule
     \multirow{2}{*}{CodeT5-small} & Fine-tuning  & 13.38  &  14.94 & 21.27 & 17.88 & 18.38 & 24.70 & 18.43 \\
    & Prompt tuning & \textbf{13.60} & \textbf{15.91} & \textbf{22.33} & \textbf{18.34} & \textbf{20.60} & \textbf{26.95} & \textbf{19.62} \\
    \midrule
     \multirow{2}{*}{CodeT5-base} & Fine-tuning & 13.70 & 15.80 & 22.60 & 17.97 & 19.56 & 25.77 & 19.23 \\
   & Prompt tuning & \textbf{14.29} & \textbf{16.04} & \textbf{23.11} & \textbf{18.52} & \textbf{19.72} & \textbf{27.06} & \textbf{19.79} \\
    \bottomrule
    \end{tabular}
    \label{tab:sum}
\end{table*}

\section{Experimental results}\label{sec:exper}



\subsection{RQ1: Effectiveness of Prompt Tuning}
In this section, we study the effectiveness of prompt tuning by comparing with the standard tuning paradigm -- fine-tuning on the three code intelligence tasks:
defect detection, code summarization, and code translation. We present the best performance achieved by our experimented prompts. Full results can be found in our project repository\footnote{{\url{https://github.com/adf1178/PT4Code}}}.
We also discuss the impact of different prompts in Section \ref{sec:RQ3}.

\textbf{Defect Detection.} 
Table~\ref{tab:defect} shows the comparison results for defect detection, in which CodeBERT and CodeT5 serve as pre-trained models. We can observe that prompt tuning always outperforms fine-tuning across different pre-trained models. For example,
prompt tuning obtains an improvement of 3.30\% over fine-tuning on CodeBERT. For CodeT5-small and CodeT5-base, the improvements are 1.51\% and 1.26\%, respectively. We also perform a statistical significance test (t-test) on defect detection task, and the results show that prompt tuning outperforms fine-tuning at the significance level at 0.05 (p-value 0.048). The results indicate that prompt tuning is more effective than fine-tuning for pre-trained models with different architecture or different sizes on the defect detection task.

\begin{figure}
    \centering
    \includegraphics[width=0.47\textwidth]{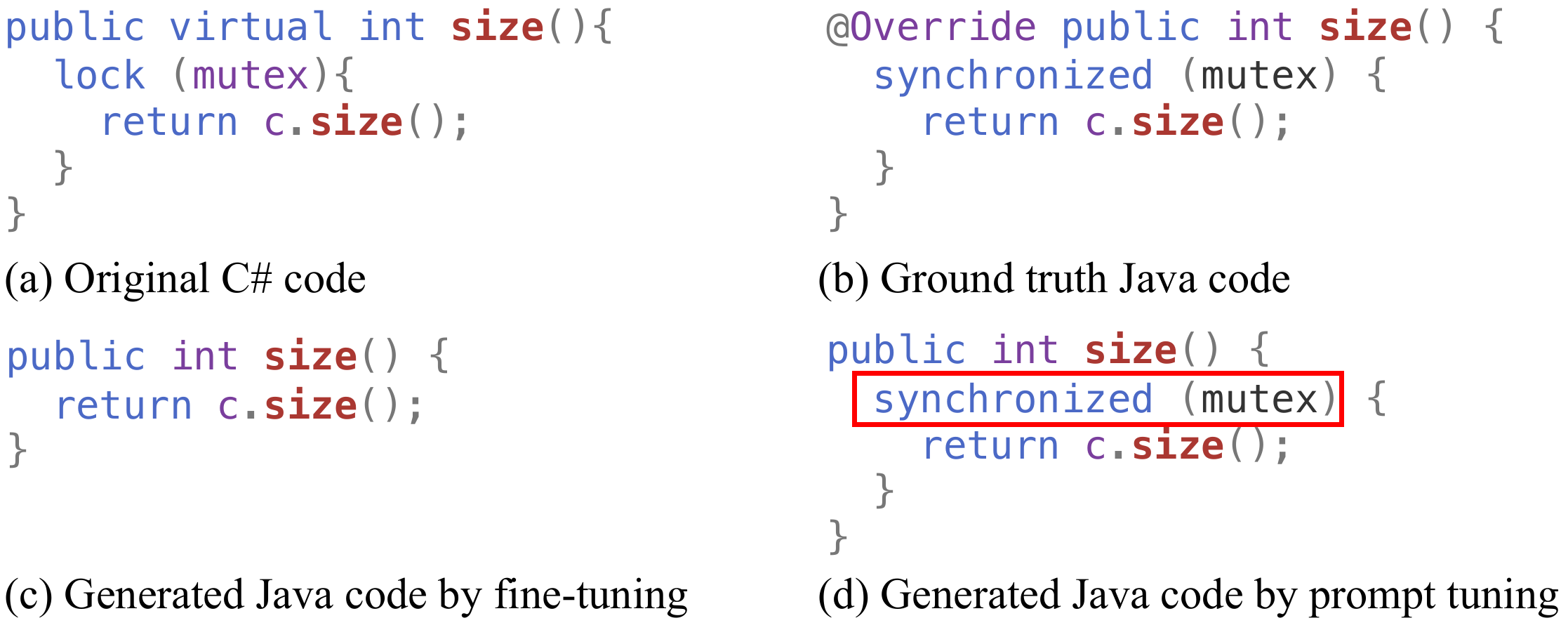}
    \caption{An example for illustrating the quality of code snippets translated by
    fine-tuning and prompt tuning in the code translation task, respectively, where the pre-trained model is CodeT5-small.}
    \label{fig:case1}
\end{figure}

\textbf{Code Summarization.} Since CodeBERT is an encoder-only model, we only involve CodeT5 as the pre-trained model on the code summarization task. Table \ref{tab:sum} presents the BLEU-4 scores achieved by prompt tuning and fine-tuning for different programming languages.
We can observe consistent improvement on overall performance as in the defect detection task: compared with fine-tuning, prompt tuning obtains an improvement of 6.46\% and 2.91\% when using CodeT5-small and CodeT5-base as pre-trained models, respectively. Looking into specific programming language, prompt tuning also always achieves better summarization performance than fine-tuning. It shows the largest advancement for the code written in PHP, with increase rate at 9.11\% and 5.01\% on CodeT5-small and CodeT5-base, respectively. Moreover, prompt tuning can perform statistically better than fine-tuning at the significance level 0.05 on code summarization with a p-value 0.019. The results indicate the effectiveness of prompt tuning in the code summarization task. 


\begin{table*}[]
    \centering
    \caption{Experimental results on code translation tasks: Java-C\# and C\#-Java. }
    \vspace{-6pt}
    \begin{tabular}{cc|ccc|ccc}
    \toprule
       \multicolumn{2}{c|}{\multirow{2}{*}{Methods}}  & \multicolumn{3}{c|}{C\# to Java} & \multicolumn{3}{c}{Java to C\#}\\
       
       &  & BLEU & Accuracy & CodeBLEU & BLEU & Accuracy & CodeBLEU \\
        \midrule
        \multirow{2}{*}{CodeT5-small} & Fine-tuning & 78.67 & 65.40 & 82.55 & 82.29 & 63.80 & 87.01 \\
        &Prompt tuning & \textbf{79.59} & \textbf{66.00} & \textbf{83.06} & \textbf{83.33} & \textbf{64.30} & \textbf{87.99} \\
        \midrule
        \multirow{2}{*}{CodeT5-base} & Fine-tuning & 79.45 & \textbf{66.10} & 83.96 & 83.61 & 65.30 & 88.32 \\
        &Prompt tuning & \textbf{79.76} & \textbf{66.10}& \textbf{84.39} & \textbf{83.99} & \textbf{65.40} & \textbf{88.74} \\
        \bottomrule
        
    \end{tabular}
    \label{tab:trans}
\end{table*}

\textbf{Code Translation.} 
For the task, we only involve the pre-trained CodeT5 model for evaluating the performance of prompt tuning. The results of prompt tuning and fine-tuning based on CodeT5 are depicted in Table \ref{tab:trans}.
From the table, we can see that prompt tuning outperforms fine-tuning in both directions of translation. 
Comparing with fine-tuning, prompt tuning achieves an average improvement of 1.22\%, 0.85\%, and  0.87\% on both directions for BLEU, Accuracy and CodeBLEU, respectively. The improvement
demonstrates the effectiveness of prompt tuning on this task. To better illustrate how prompt tuning improves the quality of code translation, we give an example in Figure \ref{fig:case1}. From the example, we can see that using fine-tuning methods, CodeT5-small does not accurately translate the C\# code into the corresponding Java version
by missing an important synchronized lock statement \textit{``synchronized (mutex)''}, while it can output more accurately with
prompt tuning. We attribute this improvement to the learned prior knowledge carried by the prefix soft prompts. Through the powerful prior knowledge, CodeT5 can quickly adapt to the translation of the code in C\# to Java,
and pay more attention to language-specific grammar details. But fine-tuning methods can only make the model learn the translation direction
after multiple iterations of training, the model may fail to focus on the important part such as ``\textit{lock}'' in the input.


Based on the performance of all the three tasks, we find that prompt tuning is more effective than fine-tuning. Another interesting observation is that the improvement of prompt tuning on CodeT5-small is 1.51\%, 6.46\%, and 1.22\%, respectively, which is higher than that on CodeT5-base, with the increase rat at 1.26\%, 2.91\%, and 0.43\%, respectively. This may be attributed to that CodeT5-base is a significantly larger model than CodeT5-small (220 million v.s. 60 million parameters), and prompt tuning (768 parameters per token). The observation suggests that prompt tuning shows more obvious improvement than fine-tuning for smaller pre-trained models.

\finding{1}{Prompt tuning is more effective than fine-tuning on the code intelligence tasks, with respect to different pre-trained models and different programming languages. Besides, the advantage of prompt tuning is more obvious for smaller pre-trained models. 
}




\begin{table*}[ht]
    \centering
    \caption{Classification accuracy (\%) on defect detection in low-resource scenario. `-' denotes the model fails to converge due to extreme lack of training data.
    }
    \begin{tabular}{cc|ccccc}
    \toprule
     \multicolumn{2}{c}{Method} & Zero shot  &  16 shots & 32 shots & 64 shots & 128 shots \\
     \midrule
     \multirow{2}{*}{CodeBERT}&Fine-tuning & 50.52 &  52.15 & 53.01 & 53.61 & 55.28  \\
     &Prompt tuning& \textbf{53.99} & \textbf{52.98} & \textbf{53.83}  & \textbf{54.28} & \textbf{56.19} \\
     \midrule
     \multirow{2}{*}{CodeT5-small}&Fine-tuning & - & - & 51.22 & 52.10 & 54.28 \\
     &Prompt tuning & - & - & \textbf{52.36} & \textbf{53.59}  & \textbf{55.04}\\
     \midrule
     \multirow{2}{*}{CodeT5-base}&Fine-tuning & - & - & 51.25  & 52.64 & 54.52\\
     &Prompt tuning & - &- & \textbf{52.44} & \textbf{53.82} & \textbf{55.47}\\
     \bottomrule
    \end{tabular}
    \label{tab:few_defect}
\end{table*}

\begin{table}[]
    \centering
    \caption{Experimental results (BLEU-4 score) on cross-language code summarization. The models are trained on Python or Java datasets, and tested on Ruby, JavaScript and Go, respectively.}
    \begin{tabular}{c|c|ccc}
    \toprule
    Training  & Methods & Ruby & JavaScript & Go \\
\midrule
\multicolumn{5}{c}{CodeT5-small} \\
\midrule
\multirow{2}{*}{Python}& Fine-tuning & 12.75 & \textbf{12.37} & 11.57 \\
   &   Prompt tuning & \textbf{13.01} & 12.35 & \textbf{12.15} \\
 \midrule
 \multirow{2}{*}{Java}  & Fine-tuning & 12.20 & 11.45 & 10.96 \\
   & Prompt tuning & \textbf{12.59} & \textbf{11.84} & \textbf{11.15} \\
      \midrule
        \midrule
        \multicolumn{5}{c}{CodeT5-base} \\
        \midrule
\multirow{2}{*}{Python}& Fine-tuning & 13.06 & 12.81 & 12.89 \\
   &   Prompt tuning & \textbf{13.37} & \textbf{13.11} & \textbf{14.27} \\ 
 \midrule
 \multirow{2}{*}{Java}  & Fine-tuning & 12.67 & 11.50 & 11.88 \\
   &  Prompt tuning & \textbf{13.13} & \textbf{11.99} & \textbf{12.96} \\
        \bottomrule
    \end{tabular}
    \label{tab:cross}
\end{table}

\begin{table}[]
    \centering
    \caption{Classification accuracy (\%) of comparing the performance of CodeBERT model on defect detection task via different prompt templates. The verbalizer is fixed as $\bm{+}$: ``\textit{bad}", ``\textit{defective}"; $\bm{-}$:``\textit{perfect}", ``\textit{clean}".
    The underlined texts are 
    replaced by virtual tokens in the corresponding vanilla soft prompt. 
    }
    \scalebox{0.9}{\begin{tabular}{cc|cc}
    \toprule
      \multirow{2}{*}{Hard Prompt} & \multirow{2}{*}{Vanilla Soft Prompt}  &  \multicolumn{2}{c}{Accuracy} \\
      & & Hard & Soft \\
      \midrule
          $[X]$ \underline{The code is} $[Z]$ & $[X]$ $[SOFT]*3$ $[Z]$ & 63.68 & 63.15 \\
      \midrule
      \underline{A} $[Z]$ \underline{code} $[X]$ & $[SOFT]$ $[X]$ $[SOFT]$ $[Z]$ & 63.36 & 62.95 \\
      \midrule  
      $[X]$ \underline{It is} $[Z]$ & $[X]$ $[SOFT][SOFT]$ $[Z]$ & 63.92 & 63.39\\
      \midrule
      \underline{The code} $[X]$ \underline{is} $[Z]$ & $[SOFT]*2$ $[X]$ $[SOFT]$ $[Z]$ & \textbf{64.17} & 63.34 \\
     \bottomrule
    \end{tabular}
    }
    \label{tab:template_defect}
\end{table}

\begin{table}[]
    \centering
    \caption{Classification accuracy (\%) 
    of different verbalizers on the defect detect task, where the pre-trained model is CodeBERT.
    The template is ``The code $[X]$ is $[Z]$''.}
    \vspace{-6pt}
    \scalebox{0.94}{\begin{tabular}{rl|c}
    \toprule
      \multicolumn{2}{c|}{Verbalizer}  & Accuracy \\
      \midrule
      $\bm{+}$: \ ``Yes"  & $\bm{-}$: \  ``No" & 63.08 \\
      \midrule
      $\bm{+}:$ \  ``bad" & $\bm{-}:$ \ ``perfect" & 63.71 \\
      \midrule
      $\bm{+}:$ \ ``bad", ``defective" & $\bm{-}:$ \ ``clean", ``perfect" & 64.17 \\
      \midrule
      \multicolumn{2}{c|}{$\bm{+}:$ \ ``bad", ``defective", ``insecure"} & \multirow{2}{*}{63.26} \\
    \multicolumn{2}{c|}{$\bm{-}:$ \ ``clean", ``perfect", ``secure" } &  \\
    \midrule
      \multicolumn{2}{c|}{$\bm{+}:$ \ ``bad", ``defective", ``insecure", ``vulnerable} & \multirow{2}{*}{63.10} \\
    \multicolumn{2}{c|}{$\bm{-}:$ \ ``clean", ``perfect", ``secure", ``invulnerable" } &  \\
      
      \bottomrule
    \end{tabular}
    }
    
    \label{tab:verb}
     \vspace{-6pt}
\end{table}


\subsection{RQ2: Capability of Prompt Tuning in Different Data Scarcity Scenarios}

\begin{figure*}
    \centering
    \includegraphics[width=0.92\textwidth]{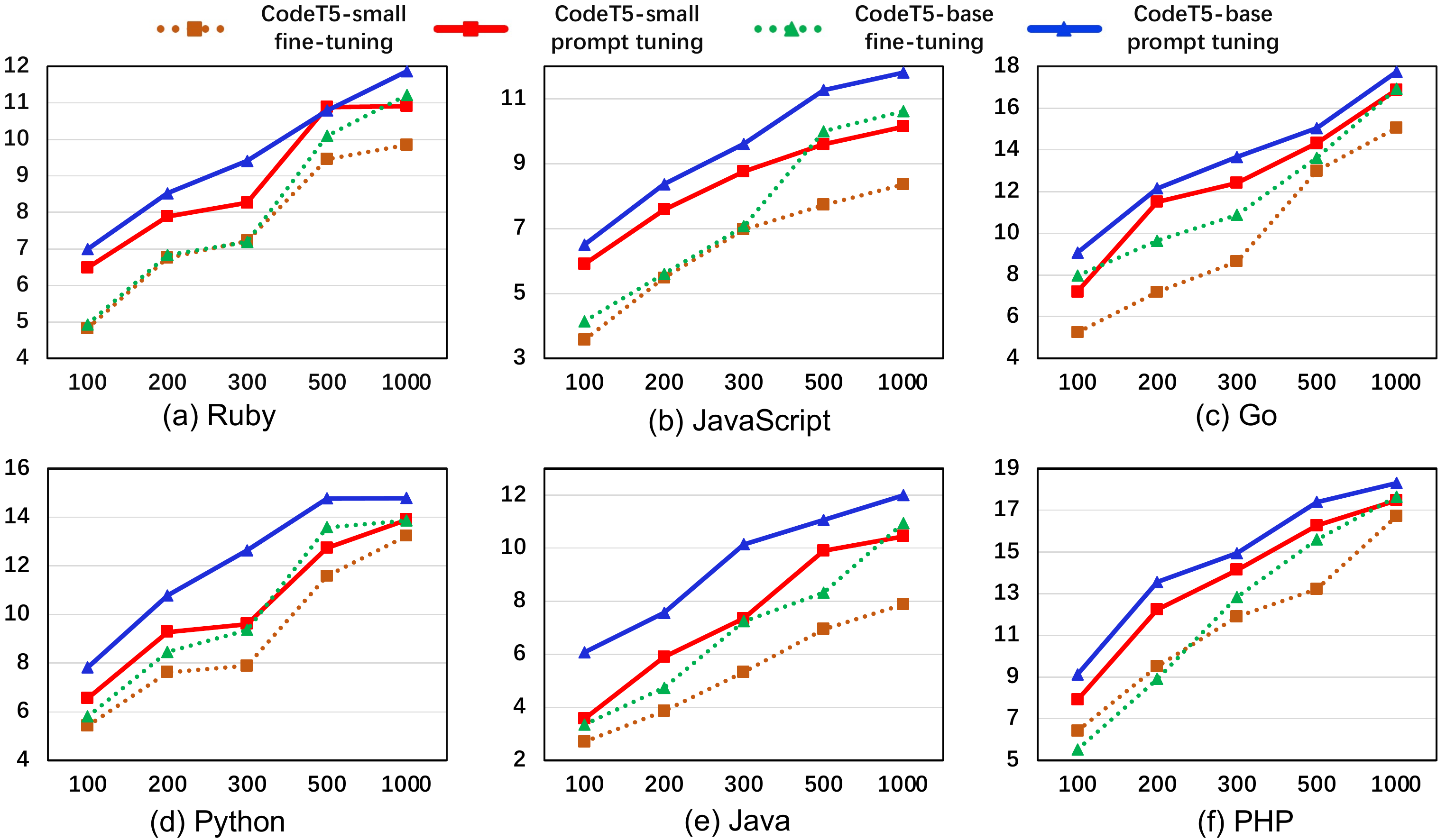}
    \caption{Results of fine-tuning and prompt tuning on code summarization task in low resource scenarios. The horizontal axis indicates the number of training instances while the vertical axis means the BLEU-4 score.} 
    \label{fig:sum_low}
\end{figure*}

Considering that the performance of fine-tuning is known to strongly rely on the amount of downstream data~\cite{fu2017easy, min2021noisy, guo2019spottune}, while scenarios with scarce data in source code are common~\cite{chai2022cross, roziere2020unsupervised,sun2022importance}. In this section, we study
how well prompt tuning can handle the data scarcity scenarios. We
focus on two kinds of data scarcity settings: 1) low-resource scenario, in which there are significantly few training instances, and 2) cross-domain scenario, in which the model is trained on a similar data-sufficient domain and tested on target domain.


\textbf{Performance in low-resource scenario.} We study the performance of prompt tuning in
low-resource setting on the classification task, i.e., defect detection, and one generation task, i.e., code summarization.
We simulate this setting by randomly select a small subset of
training instances (also called \textit{shots}) in the original dataset. To avoid randomness in data selection, we produce each subset five times with different seeds and run four times on each dataset. The average results are reported.




For the defect detection task, we choose 0, 16, 32, 64, and 128 training shots per class to create five small training subsets.
Table~\ref{tab:few_defect} presents the accuracy achieved by prompt tuning and fine-tuning regarding the five settings. Note that in zero-shot settings, no tuning data are involved. Given test data, the fine-tuning model directly generates target labels (defective or clean); while the prompt tuning model predicts the label words. By comparing the results with those in the full-data setting in Table~\ref{tab:defect}, we can find that the model performance shows severe drop. For the CodeT5 model, it even does not converge under the zero-shot and 16-shot settings due to the extreme lack of training data. The low results are reasonable since pre-trained models require task-specific data for better adapting to downstream tasks. However, we observe that with prompt tuning, the pre-trained models achieve significantly better performance than the models using fine-tuning.
On average, prompt tuning outperforms fine-tuning by 2.59\%, 2.16\%, and 2.08\% on CodeBERT, CodeT5-small and CodeT5-base, respectively. 
Note that prompt tuning under zero shot setting even outperforms prompt tuning with 32 shots and fine-tuning with 64 shots. It indicates that the knowledge of pre-trained model can be elicited by the prompt without tuning the parameters.


For the code summarization task, we choose 100, 200, 300, 500, and 1000 training shots as subsets.
Figure~\ref{fig:sum_low} shows comparison on BLEU-4 scores of prompt tuning and fine-tuning CodeT5 models on different
programming languages. We can find that although the model performance drops significantly on the subsets, prompt tuning consistently outperforms fine-tuning, showing an average improvement at
28.08\% and 26.86\% for CodeT5-small and CodeT5-base, respectively. We also observe that the improvement becomes less stark with the growth of training shots. The results demonstrate that prompt tuning is more advantageous on few training data than fine-tuning.

\textbf{Performance in cross-domain scenario.} For some programming languages, the training data are generally insufficient. As shown in Table~\ref{tab:dataset}, the data sizes of languages such as Java and Python are greatly larger than those of languages including Java-script and Ruby. Cross-domain learning is one popular solution, i.e., transferring the knowledge of similar domains with sufficient data to the target domains. We use the code summarization task for evaluating the performance of prompt tuning under cross-domain setting. Considering the adequacy of training data and domain similarity, we perform training on the programming language Java or Python, and evaluate on the language with fewer data such as Ruby, JavaScript, and Go.
Table~\ref{tab:cross} shows the cross-domain BLEU-4 scores achieved by CodeT5. We can observe that prompt tuning achieves better performance than fine-tuning for most cross-domain settings, except for the adaption from Python to JavaScript. With prompt tuning, the BLEU-4 scores of CodeT5-small and CodeT5-base are increased by 2.53\% and 5.18\% on average, respectively.

\finding{2}{Prompt tuning is more effective in low-resource scenarios than fine-tuning. The fewer training instances, the larger the improvement achieved by prompt tuning. Besides, prompt tuning also shows superior performance on the cross-domain code intelligence task.}

\begin{table*}[ht]
    \centering
    \caption{Results (BLEU-4 scores) of {prompt tuning with different prompt templates}
    on the code summarization task. There is no verbalizer for the prompts of generation tasks.
    }
    \scalebox{0.95}{\begin{tabular}{cc|cccccc|c}
    \toprule
   \multicolumn{2}{c|}{$f_{prompt}(\cdot)$}      & Ruby & JavaScript & Go & Python & Java & PHP & Overall \\

    \midrule
    \multirow{5}{*}{CodeT5-small} &\underline{Summarize $[LANG]$} $[X]$ $[Z]$& 13.45 & 15.01 & 21.20 & 17.82 & 18.43 & 24.52 & 18.41 \\
    & $[SOFT]*2$ $[X]$ $[Z]$& 13.33 & 14.96 & 21.17 & 17.93 & 18.29 & 24.61 & 18.38 \\
    \specialrule{0em}{1pt}{1pt}
    &\underline{Generate comments for $[LANG]$} $[X]$ $[Z]$& 13.44 & 14.96 & 21.24 & 17.90 & 18.52 & 24.46 & 18.42 \\
   & $[SOFT]*4$ $[X]$ $[Z]$& 13.49 & 14.87 & 21.29 & 17.92 & 18.34 & 24.68 & 18.44 \\
    \midrule
    \multirow{5}{*}{CodeT5-base}&\underline{Summarize $[LANG]$} $[X]$ $[Z]$& 13.67 & 15.91 & 22.51 & 18.00 & 19.63 & 25.76 & 19.25 \\
    &$[SOFT]*2$ $[X]$ $[Z]$& 13.86 & 15.75 & 22.48 & 18.12 & 19.52 & 25.91 & 19.27 \\
   \specialrule{0em}{1pt}{1pt}
    &\underline{Generate comments for $[LANG]$} $[X]$ $[Z]$& 13.68 & 15.84 & 22.49 & 18.03 & 19.59 & 25.88 & 19.25 \\
    & $[SOFT]*4$ $[X]$ $[Z]$& 13.74 & 15.82 & 22.63 & 18.06 & 19.60 & 25.83 & 19.28 \\
    \bottomrule
    \end{tabular}
    }
    \label{tab:template_sum}
\end{table*}
\begin{figure*}
    \centering
    \includegraphics[width=0.95\textwidth]{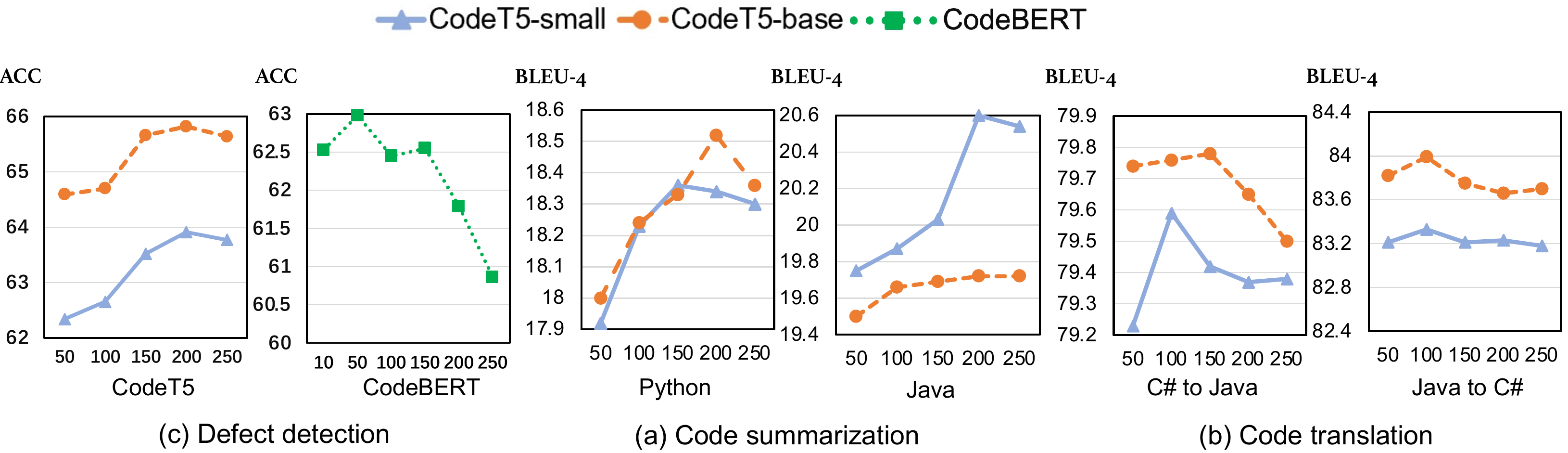}
    \caption{BLEU-4 score of comparing the performance of CodeT5 models on code summarization and code translation with different prefix lengths. The horizontal axis indicates the length of prefix. }
    \label{fig:prefix_sum}
\end{figure*}

\subsection{RQ3: Impact of Different Prompts}
\label{sec:RQ3}

In this RQ, we explore the impact of different prompts on the performance of prompt tuning. We focus on the following three aspects: 1) hard prompt template; 2) hard prompt v.s. vanilla soft prompt; and 3) length of prefix soft prompt. 


\subsubsection{Different Hard Prompt Templates.} 
There are two factors that can impact the performance of hard prompts, including the template design and verbalizer. Due to the space limit, we present the evaluation results on the classification task, i.e., defect detection, and one generation task, i.e., the code summarization. Note that we have the same observation for the code translation task.

\textbf{Template Design.} The natural language tokens in hard prompt templates are manually defined. To study the impact of different tokens in the template, we conduct experiments with fixed verbalizers. Table~\ref{tab:template_defect} and Table~\ref{tab:template_sum} show the results on the defect detection task and code summarization task, respectively. Comparing the row 2-5 in Table~\ref{tab:template_defect}, we can find that the template design impacts the model performance.
For example, when using the hard prompt $``The\ code\ [X]\ is\ [Z] "$, CodeBERT outperforms the case when using $``A\ [Z]\ code\ [X]"$ by 1.39\%. 
In addition, by changing $``The\ code\ [X]\ is\ [Z] "$ to $``[X]\ The\ code\ is\ [Z] "$ in which only the token order is different,
a drop in performance by 0.8\% is observed. However, comparing row 2 and 4 in Table~\ref{tab:template_sum}, we can find that the model performance is less affected by the template design for the code summarization task. This may be attributed to that only few prompt tokens in the templates
can hardly provide helpful guidance for the large solution space in the code summarization task. Thus, we achieve that the template design for hard prompt is more important for the classification task than the generation task.


\textbf{Different Verbalizers:}
We
fix the hard prompt template as $\ \ $ `` $The\ code\ [X]\ is\ [Z]$" and analyze the impact of different verbalizers on the model performance. Specifically, we choose task-relevant label words for the verbalizers, with the results on the defect prediction task shown in Table~\ref{tab:verb}. We can observe that different verbalizers influence the performance of prompt tuning. When choosing label words such as \textit{``yes''} and \textit{``no''} (row 2) rather than adjectives to fill the answer slot $[Z]$, the result is 0.99\% lower than that of using adjectives in the verbalizer (row 3). It indicates that constructing verbalizer with correct grammar is helpful for the prediction. Comparing row 3-6, we can also find that increasing the number of label words is not always beneficial for the model performance, which may be because more label words could introduce bias to the prediction results. When using two label words for indicating each class, the model presents the highest performance.

\subsubsection{Hard Prompt vs. Vanilla Soft Prompt.}
As introduced in Section~\ref{sec:soft}, the vanilla soft prompt replaces the natural language tokens in hard prompt with virtual tokens. The comparison results on the defect detection task are illustrated in Table~\ref{tab:template_defect}. We experiment with different hard prompts, shown in the first column, with the corresponding vanilla soft prompts at the second column. From the results listed as the last two columns, we can find that hard prompts present better prediction accuracy than the corresponding vanilla soft prompts.
For the code summarization task, the results
are shown in Table~\ref{tab:template_sum}. Comparing the performance of hard prompts such as ``$Summarize \ [LANG]\ [X] [Z]$'' and ``\textit{Generate comments for} $[LANG]$ $[X]$ $[Z]$'' with the corresponding vanilla soft version,
we can observe that the difference is marginal, which may be due to the large generation space of the task. Thus, we summarize that hard prompts may be more effective than the corresponding vanilla soft prompts for classification tasks, and the advantage tends to be weakened for generation tasks.


\subsubsection{Different Lengths of Prefix Soft Prompts.} 
We also study the impact of different lengths of prefix soft prompts. We illustrate the performance 
under different prefix prompt lengths for the three tasks in Figure~\ref{fig:prefix_sum}.
As can be seen, too short or long lengths of prefix prompts can degrade the model performance. For all the tasks, prompt tuning achieves the best or nearly best performance when the length of prefix prompt set to a value between 100 and 200. In our work, the prefix lengths are set as 200, 200, and 100 for defect detection, code summarization and code translation tasks, respectively.

\finding{3}{
Prompt templates have large impact on the performance of prompt tuning. It is crucial to construct prompt templates with suitable template design
and verbalizers based on domain knowledge.
When using the prefix prompts, 
the length of prompts {has 
impact on the model performance.} 
} 

\section{Discussion}\label{sec:Discussion}

\begin{figure}[t]
    \centering
    \includegraphics[width=0.45\textwidth]{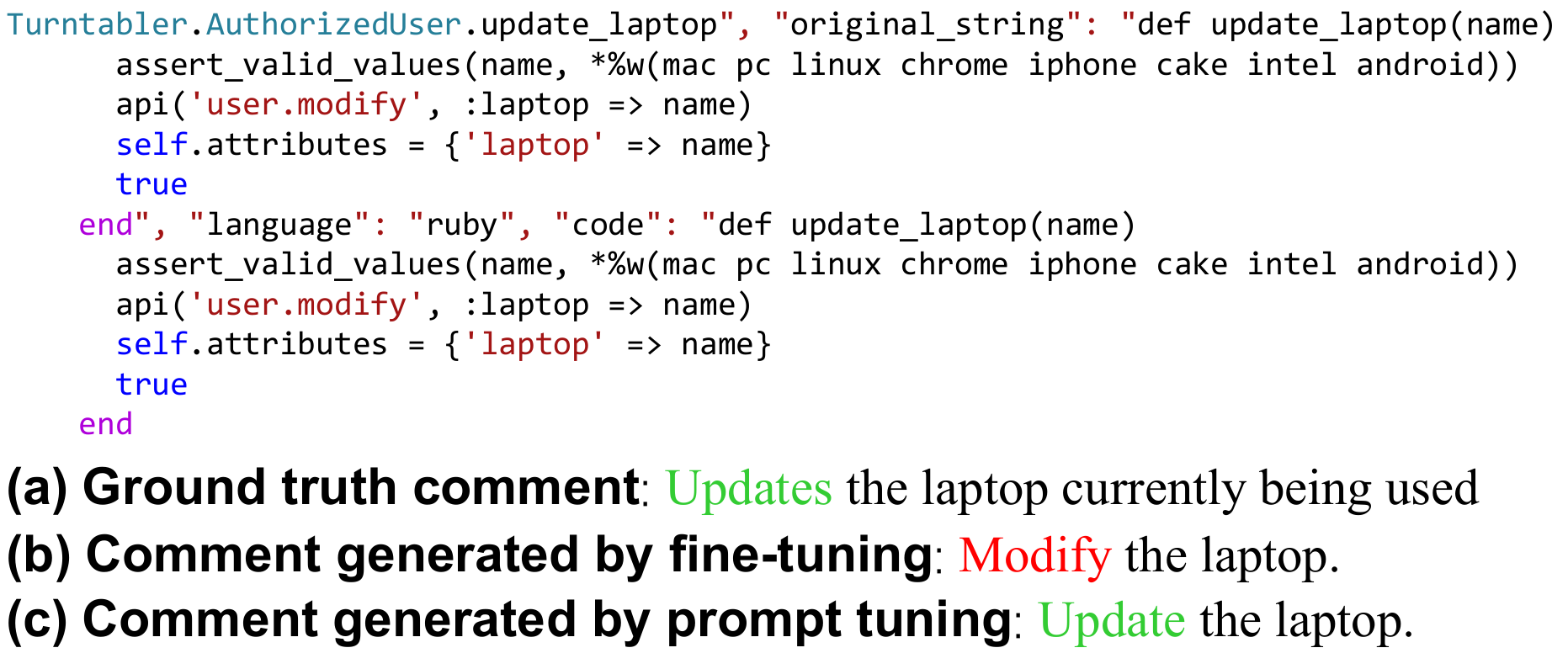}
    \caption{Case study on the code summarization task, where the pre-trained model is CodeT5-small.}
    \label{fig:case2}
\end{figure}
\subsection{Implications of Findings}\label{sec:Related Work}


\paragraph{Implication on the Utilization of Pre-trained Models} 
Prompt tuning performs well in adapting pre-trained models on code intelligence tasks. 
We observe that prompt tuning can consistently outperform fine-tuning in our experiments under full-data settings, data scarcity settings, and cross-domain settings. The advantage of prompt tuning is especially outstanding in data scarcity settings, 
which suggests that prompt-tuning is a superior solution when there is a lack of task-specific data. 

\paragraph{Implication on the Utilization of Prompts} 
Our experiments demonstrate that different templates and {verbalizers} influence the performance of the code intelligence tasks. The templates that have the same semantics but different prompt tokens
can lead to different performance results. 
Researchers could try different combinations of the words in their templates and evaluate the effectiveness through experiments.
Besides, although the vanilla soft prompt is helpful to reduce the manual cost of  prompt template designing, the best performance is achieved mostly by well-designed hard prompt.
Furthermore, we find that the performance of prefix soft prompt varies
with its length. Determining the best length of the prompt for a downstream task is difficult. 
Based on our experiments, in general, 
promising results can be achieved by soft prompt when the length is between 100 and 200.


\begin{figure}[t]
    \centering
    \includegraphics[width=0.45\textwidth]{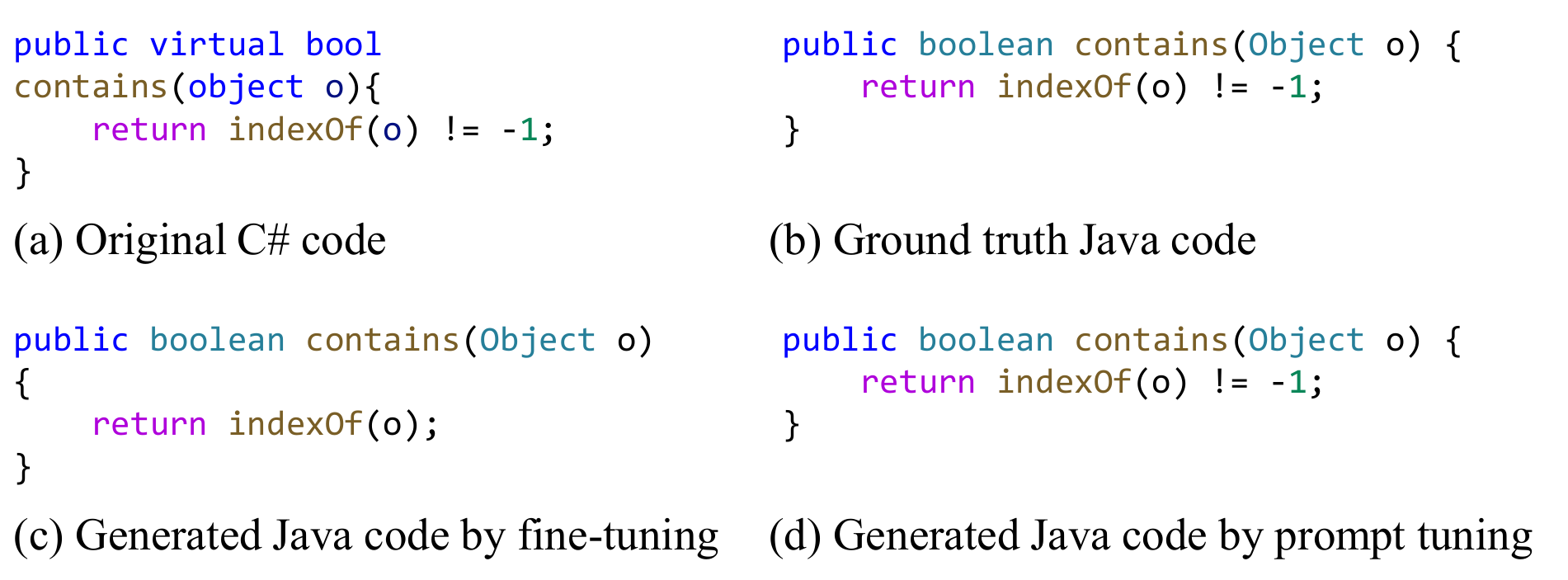}
    \caption{Case study on the code translation task, where the pre-trained model is CodeT5-small.}
    \label{fig:case3}
\end{figure}
\subsection{Case Study}
In this section, we provide additional case studies to qualitatively compare prompt tuning with fine-tuning.

The case in Figure~\ref{fig:case2} shows a Ruby code snippet with comments generated by fine-tuning and prompt tuning models. From the case we can observe that the fine-tuning model is mislead by the word ``\textit{modify}" in the code snippet and fails to capture the main functionality ``\textit{update}". Quite the opposite, the prompt tuning model accurately summarizes the code snippet.

We also give another case in code translation task in Figure~\ref{fig:case3}. The original C\# code (a) is to check whether object \textit{o} is contained. The code translated by fine-tuning model (c) only returns the index of \textit{o} but does not compare it with -1, where the code semantic changes. However, the prompt tuning model generates the identical Java code (d) with the ground truth one (b).


\subsection{Future Directions}

Based on the findings and implications, we suggest two possible future directions for prompt tuning on source code. First, we suggest future research to consider more  characteristics of source code, like syntactic structures, in the design of template and the choices of verbalizer. Experiment results demonstrate that domain knowledge plays an important role on the design of prompts. As code structure information has been demonstrated on the design of regular DL models for code-related tasks \cite{guo2020graphcodebert, zhang2019novel, liu2020atom, liu2020retrieval, gu2021cradle}, we believe that the domain knowledge carried by them can also help the design of prompts.
Second, through constructing cloze-style prompt template, the factual knowledge and biases contained in the pre-trained models can be investigated \cite{zhong2021factual, petroni2019language, jiang2020can}. 
Researchers can focus on improving the interpretability and robustness of pre-trained models and designing novel pre-training tasks in the future.

\subsection{Threats To Validity}
We have identified the following major threats to validity:

\textbf{Limited datasets.} The experiment results are based on a limited number datasets for each code intelligence task.
The selection of data and datasets may bring bias to the results. To mitigate this issue, we choose the most widely-used datasets for each code-related task, modify the seeds and run the sampling multiple times. 
We also plan to collect more datasets
in the future to better evaluate the effectiveness of prompt tuning.



\textbf{Limited downstream tasks.} Our experiments are conducted on three code intelligence tasks, including one classification task and two generation tasks. Although these tasks are the representative ones in code intelligence,
there are many other tasks, such as code search~\cite{gu2018deep,cambronero2019deep} and bug fixing~\cite{zhou2012should, mastropaolo2021studying}.  
We believe that we could obtain similar observations on these tasks since they can all be formulated as either classification tasks or generation tasks for source code. We will evaluate more tasks with prompt tuning in our future work.

\textbf{Suboptimal prompt design.} We demonstrate that prompt tuning can improve the performance of pre-trained models. However, the prompts we use in this paper may not be the best ones. 
It is challenging to design the best prompt templates and verbalizers, which will be an interesting future work.


\section{Related Work}\label{sec:Related Work}

\subsection{Pre-training on Programming language}
Code intelligence aims at learning the semantics of programs to facilitate various program comprehension tasks, such as code search, code summarization, and bug detection~\cite{iyer2016summarizing,zhang2019novel,gu2018deep,wang2020modular, leclair2020improved,wan2018improving,li2019improving, lam2017bug}.
 Recently, inspired by the huge success of pre-trained models in NLP, a boom of pre-training models on programming languages arises. CuBERT \cite{kanade2020learning} and CodeBERT \cite{feng2020codebert} are two pioneer works. CuBERT utilizes the 
MLM pre-training objective in BERT \cite{devlin2018bert} to obtain better representation of source codes. CodeBERT is able to learn NL-PL representation via replaced token detection task \cite{clark2019electra}. Svyatkovskiy et al.  \cite{svyatkovskiy2020intellicode} and Kim et al. \cite{kim2021code} train GPT-2 \cite{radford2019language} on large scale programming languages for solving code completion task. The work GraphCodeBERT \cite{guo2020graphcodebert} leverages data flow graph (DFG) in model pre-training stage, making model better understand the code structure.

Apart from aforementioned encoder or decoder only models, pre-trained models that utilize both 
encoder and decoder 
 are also proposed  for programming languages. For example, Ahmad et al. propose PLBART \cite{ahmad2021unified}, which is able to support both understanding and generation tasks. The work  \cite{mastropaolo2021studying, elnaggar2021codetrans} utilizes text to text transfer transformer (T5) framework to solve code-related tasks. Wang et al. modify the pre-training and finetuning stages of T5 and propose CodeT5 \cite{wang2021codet5}. 

\subsection{Prompt Tuning}
The concept of prompt tuning is formed gradually. In the work \cite{petroni2019language}, the authors find that the pre-trained language models have ability to learn the factual knowledge due to the mask-and-predict pre-training approach. Therefore, pre-trained language models can be regarded as a kind of knowledge base. To measure the capability of pre-trained models to capture factual information, they propose a language model analysis dataset (LAMA). Later, Jiang et al. attempt to more accurately estimate the knowledge constrained in the language model \cite{jiang2020can}. They propose LPAQA to automatically discovery better prompt templates. 
Several works focus on exploring good templates. Yuan et al. \cite{yuan2021bartscore} replace phases in the template via a thesaurus. The work \cite{haviv2021bertese} utilizes a neural prompt rewriter to improve the model performance. 
Aforementioned works explore the manual templates or hard templates (meaning the words in the template are fixed and not learnable). Researchers also attempt to optimize the template in the training process (soft prompt) \cite{li2021prefix, zhong2021factual, tsimpoukelli2021multimodal}. For example, Li et al. add an additional learnable matrix in front of the input embedding \cite{li2021prefix}. Zhong et al. propose to initialize these matrices by natural language tokens for more effective optimization \cite{zhong2021factual}.   
Recently, a series of works also study prompts in pre-training stage. They find that the behavior of language models can be manipulated to predict desired outputs \cite{radford2018improving, petroni2020context, brown2020language, schick2020s}, sometimes even require no task specific training. 
In our work, we adapt prompt tuning in code intelligence tasks to exploit knowledge about both natural language and programming languages captured by pre-trained models. 

\section{Conclusion}\label{sec:conclusion}
In this paper, we experimentally investigate the effectiveness of prompt tuning on three code intelligence tasks with two pre-trained models. Our study shows that prompt tuning can outperform fine-tuning 
under full-data settings, data scarcity settings, and cross-domain settings. We summarize our findings and provide implications that can help researchers exploit prompt tuning effectively in their code intelligence tasks. 
Our source code and experimental data are publicly available at:\textit{\url{https://github.com/adf1178/PT4Code}}.


\normalem
\bibliographystyle{ACM-Reference-Format}
\bibliography{sample-base}

\end{document}